\documentclass[aps,prl,preprint,superscriptaddress]{revtex4}   
\usepackage{graphicx}
 
\def\ket#1{\left|#1\right>}
\newcommand{\be}{\begin{equation}}
\newcommand{\ee}{\end{equation}}
\newcommand{\bea}{\begin{eqnarray}}
\newcommand{\eea}{\end{eqnarray}}

\begin{document}

\title{Towards high-speed optical quantum memories }

\author{K. F. Reim}

\affiliation{Clarendon Laboratory, University of Oxford, Parks Road, Oxford OX1 3PU, UK}

\author{J. Nunn}
\affiliation{Clarendon Laboratory, University of Oxford, Parks Road, Oxford OX1 3PU, UK}

\author{V. O. Lorenz}
\affiliation{Department of Physics, University of Delaware, Newark, DE 19716, USA}
\affiliation{Clarendon Laboratory, University of Oxford, Parks Road, Oxford OX1 3PU, UK}

\author{B. J. Sussman}
\affiliation{Clarendon Laboratory, University of Oxford, Parks Road, Oxford OX1 3PU, UK}
\affiliation{National Research Council of Canada, Ottawa, Ontario K1A 0R6, Canada}

\author{K.C. Lee}
\affiliation{Clarendon Laboratory, University of Oxford, Parks Road, Oxford OX1 3PU, UK}

\author{N. K. Langford}
\affiliation{Clarendon Laboratory, University of Oxford, Parks Road, Oxford OX1 3PU, UK}

\author{D. Jaksch}
\affiliation{Clarendon Laboratory, University of Oxford, Parks Road, Oxford OX1 3PU, UK}

\author{I. A. Walmsley}
\email[]{i.walmsley1@physics.ox.ac.uk}
\affiliation{Clarendon Laboratory, University of Oxford, Parks Road, Oxford OX1 3PU, UK}

\date{\today}

\begin{abstract}
Quantum memories, capable of controllably  storing and releasing a photon, are a crucial component for quantum computers \cite{Kok_2007_RMP_Linear-optical-quantum-computi} and quantum communications \cite{Duan_2001_N_Long-distance-quantum-communic}. 
So far, quantum memories \cite{Hetet_2008_PRL_Electro-Optic-Quantum-Memory-f,Riedmatten_2008_N_A-solid-state-light-matter-int,Schnorrberger_2009_PRL_Electromagnetically-Induced-Tr, Boozer_2007_PRL_Reversible-State-Transfer-betw} have operated with bandwidths that limit data rates  to MHz. Here we report the coherent storage and retrieval of sub-nanosecond low intensity light pulses with spectral bandwidths exceeding 1~GHz in cesium vapor. The novel memory interaction takes place via a far off-resonant two-photon transition in which the memory bandwidth is dynamically generated by a strong control field \cite{Gorshkov_2007_PRAMOP_Photon-storage-in-Lambda-type-,Nunn_2007_PRAMOP_Mapping-broadband-single-photo}. 
This allows for an increase in data rates by a factor of almost 1000 compared to existing quantum memories. The memory works with a total efficiency of 15\% and its coherence is demonstrated by directly interfering the stored and retrieved pulses. Coherence times in hot atomic vapors are on the order of microseconds \cite{Camacho_2009_P_Four-wave-mixing-stopped-light} --- the expected storage time limit for this memory.

\end{abstract}

\maketitle

Photons are ideal carriers of quantum information: they have a very  large potential information capacity, and do not interact with one  another, so that information coded in them is robust.  Recent  
developments in sources, detectors, gates and protocols have laid the  
ground for the construction of large-scale photonic quantum computers  
with unique capabilities \cite{Kok_2007_RMP_Linear-optical-quantum-computi,Shor_1997_SJSSC_Polynomial-Time-Algorithms-for}, as well as inter-continental quantum networks that are immune to eavesdropping~\cite{Sangouard_2009_EA0_Quantum-repeaters-based-on-ato}. However, the effects of photon loss and the  
inherently probabilistic character of some of these functions demand
the ability to store photons. The difficulty that photonic networks  
will only produce the desired results rarely is overcome if photon  
storage is possible, since this allows complex protocols to be orchestrated by holding the output of successful computations until all operations have been correctly executed \cite{Barrett_2008__Scalable-quantum-computing-wit}. Quantum memories are therefore an active area of research, with great interest focussed on reversibly mapping photons into collective atomic excitations \cite{Choi_2008_N_Mapping-photonic-entanglement-,Schnorrberger_2009_PRL_Electromagnetically-Induced-Tr}.

Key characteristics for quantum memories are long storage time, high memory efficiency, the ability to store multiple modes (i.e. multiple distinct photons) \cite{Nunn_2008_PRL_Multimode-Memories-in-Atomic-E,Simon_2007_PRL_Quantum-Repeaters-with-Photon-} and high bandwidth. High bandwidth allows the storage of temporally short photons, enabling quantum information to be processed at a higher ``clock rate'', increasing the number of computational cycles that can be completed before decoherence sets in. This can be difficult to achieve with atomic memories, since photons must be stored in long-lived atomic states with narrow linewidths. Here we demonstrate the storage of signal pulses with a bandwidth $7000$ times larger than the natural width of the cesium $D2$ line that mediates the interaction.

Previously implemented memory protocols include electromagnetically induced transparency (EIT), controlled reversible inhomogeneous broadening (CRIB) and atomic frequency combs (AFC).
EIT based memories \cite{Eisaman_2005_N_Electromagnetically-induced-tr,Harris_1997_PT_Electromagnetically-Induced-Tr,Liu_2001_N_Observation-of-coherent-optica}  utilize the extreme dispersion of an induced transparency window to modify the group velocity, and controllably stop, store and retrieve light pulses.
CRIB \cite{Alexander_2006_PRL_Photon-Echoes-Produced-by-Swit,Kraus_2006_PRAMOP_Quantum-memory-for-nonstationa,Staudt_2007_PRL_Fidelity-of-an-Optical-Memory-} is a photon echo technique that uses an artificial inhomogeneous broadening of the atomic resonance. Reversing this broadening during the readout process causes the atomic spins to rephase and collectively reemit the original signal. 
In the AFC protocol \cite{Riedmatten_2008_N_A-solid-state-light-matter-int,Chaneliere_2009_0_Efficient-light-storage-in-a-c} an artificially created atomic frequency comb absorbs the incident signal, and the periodic structure of the absorption spectrum results in a subsequent re-phasing and re-emission of the stored signal. These protocols are resonant; off-resonant light storage has also been implemented via four-wave mixing \cite{Camacho_2009_P_Four-wave-mixing-stopped-light}, stimulated Brillouin scattering \cite{Zhu_2007_S_Stored-light-in-an-optical-fib} and via the gradient echo memory (GEM) protocol \cite{Hosseini_2009_N_Coherent-optical-pulse-sequenc}.  For all these protocols typical storage times range from $\mu$s to  ms and  achieved efficiencies from 1\% to 15\%, although two experiments have reported higher values for either storage time or efficiency \cite{Novikova_2008_PRAMOP_Optimal-light-storage-with-ful, Longdell_2005_PRL_Stopped-Light-with-Storage-Tim}. The reported bandwidths range from kHz to MHz.

In this letter  we present the experimental demonstration of a coherent, efficient and  broadband Raman memory for light. In a Raman memory, the bandwidth is generated dynamically by ancillary write/read pulses, which dress the narrow atomic resonances to produce a broad virtual state to which the signal field couples. The off-resonant nature of the scheme confers some appealing features \cite{Kozhekin_2000_PR_Quantum-memory-for-light,Nunn_2007_PRAMOP_Mapping-broadband-single-photo,Gorshkov_2007_PRAMOP_Photon-storage-in-Lambda-type-}. These include \emph{(i)} as mentioned above, the ability to store broadband pulses --- the large detuning guarantees that the atomic polarization adiabatically follows the pulse envelopes, even when they are temporally short; \emph{(ii)} insensitivity to inhomogeneous broadening --- the Raman transition is detuned far beyond the Doppler linewidth of the cesium vapor, and \emph{(iii)} the property that any unstored light is transmitted without attenuation. This last feature is useful since the partial storage of a single photon entangles the memory with the optical mode of the transmitted signal. Such light-matter entanglement operations are primitives for the construction of quantum repeaters \cite{Duan_2001_N_Long-distance-quantum-communic, Sangouard_2009_EA0_Quantum-repeaters-based-on-ato}, and for enabling coherent logical information storage. In our current experiment, we used signal pulses containing several thousand photons. However, because the memory interaction is linear and coherent, the Raman protocol is a genuine quantum memory that would also work in the single photon regime. 

\begin{figure}[h]
\begin{center}
\includegraphics[width =11cm]{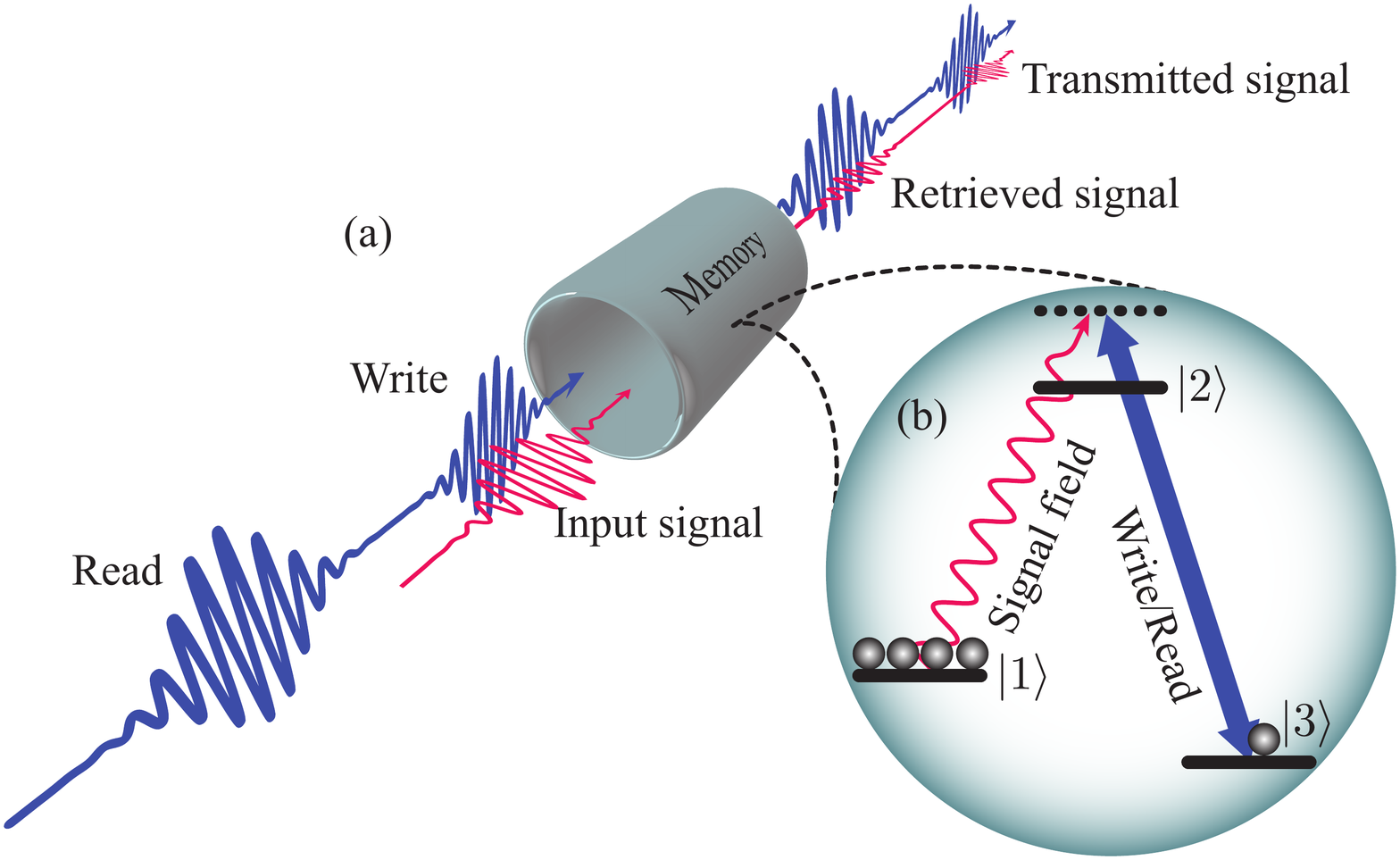}
\caption{(a) A Raman memory. The signal is directed into the memory along with a bright write pulse and is stored. If the storage is partial, any unstored signal is transmitted through the memory. A subsequent read pulse extracts the stored excitation, which emerges along with the transmitted read pulse. (b) The $\Lambda$-level structure of the atoms in the memory. The atoms are prepared in the ground state $\ket{1}$ by optical pumping. The signal is tuned into two-photon resonance with the write field; both are detuned from the excited state $\ket{2}$. Absorption of a signal photon transfers an atom from $\ket{1}$ into the storage state $\ket{3}$ via Raman scattering stimulated by the write field. Upon retrieval the interaction is reversed.}
\label{fig:memory}
\end{center}
\end{figure}

In the experiment a strong write pulse and a weak signal pulse, both broadband, are spatially and temporally overlapped and sent together into a cesium vapor cell where the Raman interaction with the storage medium takes place (see Fig.~\ref{fig:memory}). The signal pulse is mapped via a two-photon transition with the write pulse into a collective atomic excitation called a \emph{spin~wave}. At a later time a strong read pulse is sent into the vapor cell and converts the spin wave into an optical output signal that is measured on a fast detector. The $F=3,4$ `clock states' of the ground-level hyperfine manifold serve as the states $\ket{1}$ and $\ket{3}$, which are connected to the excited state $\ket{2}$ (the $6^2P_{3/2}$ manifold) via the $D2$ line at $852$~nm (supplementary information). The cesium is heated to $62.5$~$^\circ C$, so that the resonant optical depth $d\approx 1800$ associated with this transition is high. 

\begin{figure}[h]
\begin{center}
\centerline{\includegraphics[width=11cm]{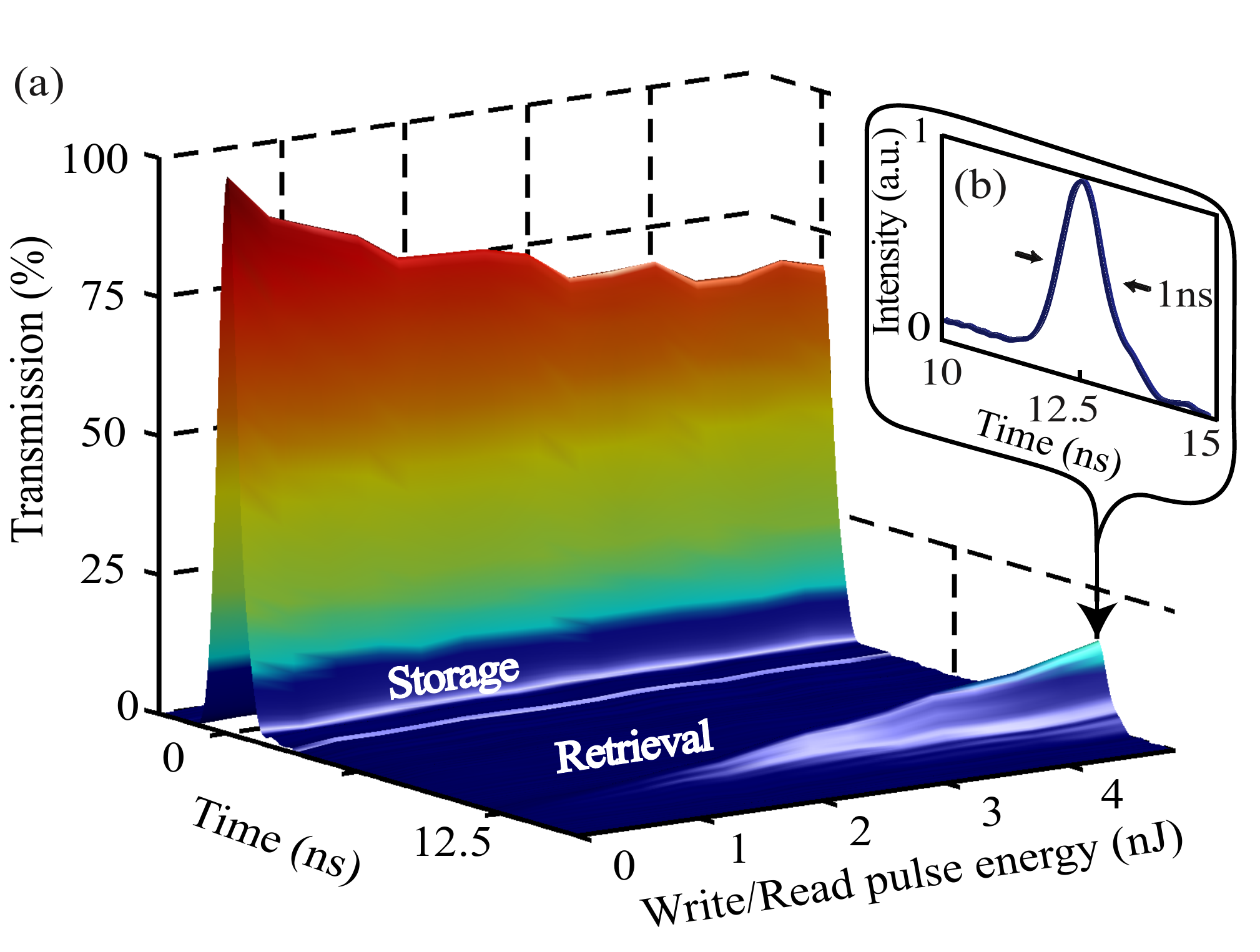}}
\caption{(a) Storage ($t=0$ ns) and retrieval ($t=12.5$ ns) of light pulses vs. write/read pulse energy. With no write pulse present (\emph{i.e.} 0 nJ), there is 100\% transmission; with the highest write/read pulse energies (4.8 nJ) this drops to 70\%, indicating that 30\% of the incident signal is stored. At $t=12.5$ ns  50\% of the stored information is retrieved giving a total memory efficiency of $15\%$. (b) Zoom of retrieved signal field showing the measured full width at half maximum (FWHM) temporal duration of 1 ns, limited by the detector response time. This shows that the bandwidth of the retrieved signal exceeds 1~GHz.}
\label{fig:abs_eff_2d}
\end{center}
\end{figure}

Experimental data for the storage and retrieval processes are displayed in figure \ref{fig:abs_eff_2d}. The storage of a signal pulse takes place at time $t=0$ and the retrieval of the stored information is carried out $12.5$ ns later. 
The storage and retrieval efficiencies depend on the write and read pulse energy; if they are zero, 100\% of the incident signal field is transmitted --- this contrasts with resonant storage protocols, in which the memory becomes absorbing when `inactive'.
Increasing the write and read pulse energy decreases the transmitted fraction of the incident signal and increases the retrieved signal. The inset in figure \ref{fig:abs_eff_2d} clearly shows the short pulse duration of the retrieved signal. The measurement is limited by the response time of the detector, which is 1~ns, corresponding to a bandwidth of 1~GHz. In practice, our Raman memory should be operating at the full control field bandwidth of 1.5 GHz (see methods). 

The time-bandwidth product $N$ of a memory quantifies the number of distinct time bins available for computational operations in a hypothetical quantum processor using the memory. We expect that the storage time of the memory is limited to several hundred $\mu$s --- typical of warm alkali vapors \cite{Camacho_2009_P_Four-wave-mixing-stopped-light} --- so that time-bandwidth products as high as $N\sim 10^5$ should be achievable.

\begin{figure}[h]
\begin{center}
\centerline{\includegraphics[width=9cm]{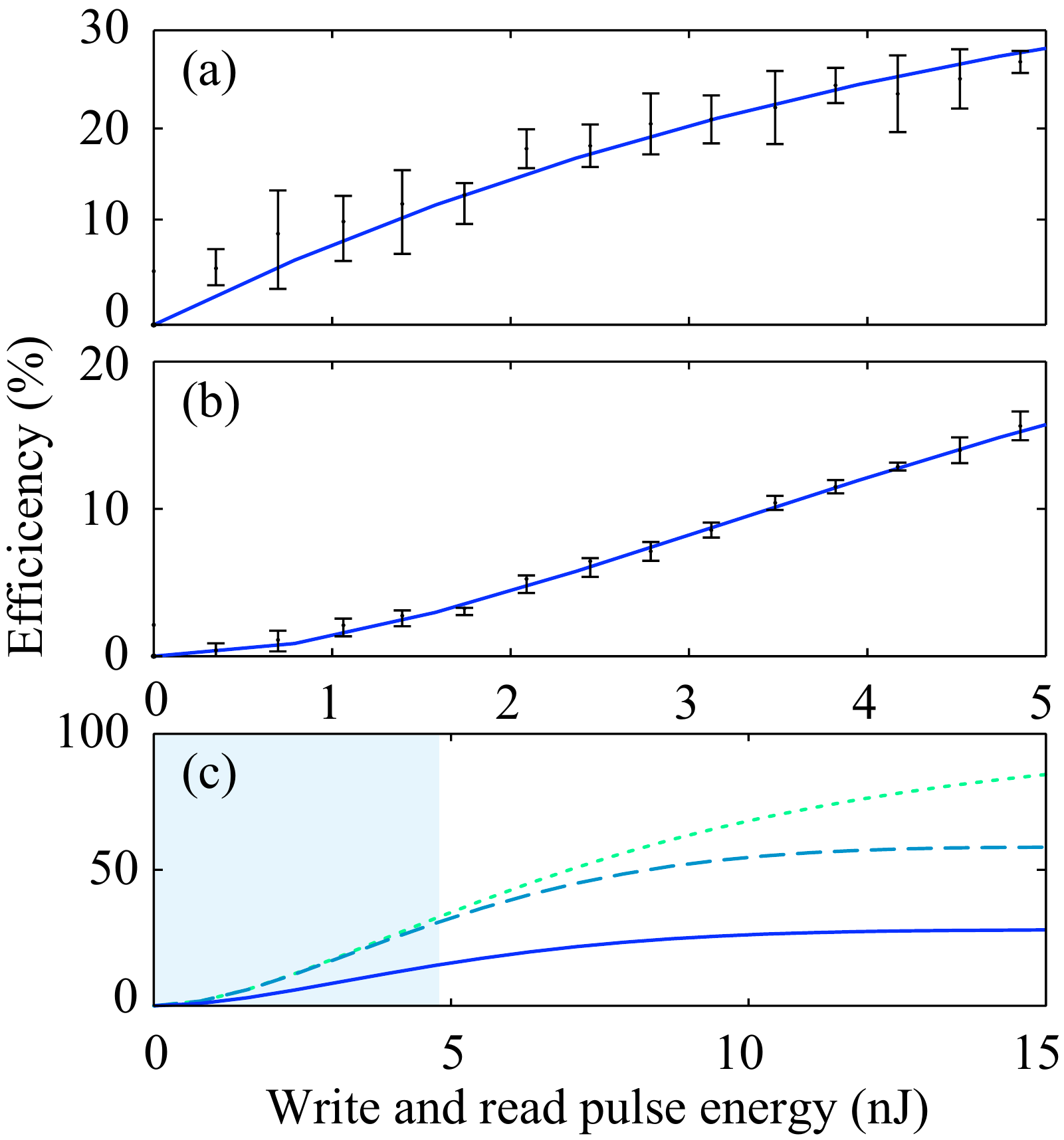}}
\caption{Dependence of memory efficiency on write/read pulse energy. (a) Storage efficiency. (b) Total efficiency. Dots and error bars indicate experimental data --- solid lines represent predicted theory. 
(c) Theoretical prediction for total efficiency extrapolated to higher pulse energies. Solid line: Efficiency for current experimental configuration with forward readout.  Dashed line: Optimal efficiency using forward retrieval, limited by re-absorption \cite{Surmacz_2008_PRAMOP_Efficient-spatially-resolved-m,Gorshkov_2007_PRAMOP_Photon-storage-in-Lambda-type-}. Dotted line: Optimal efficiency using phasematched backward retrieval. The shaded area denotes the range of pulse energies accessible with the present experiment.}
\label{fig:eff_error}
\end{center}
\end{figure}

Parts (a) and (b) of figure~\ref{fig:eff_error} show a comparison of the measured efficiencies for storage and retrieval with the predictions of a theoretical model \cite{Nunn_2007_PRAMOP_Mapping-broadband-single-photo} (supplementary information). Precise measurement of the pulse shapes requires greater temporal resolution than the current detector provides, so to apply the theory we assume Gaussian temporal profiles for all pulses; the timing and duration of the signal pulse are then adjusted to account for the dispersive effects of the etalons used to spectrally filter the signal. The observations stand in good agreement with the theory.


The retrieval efficiency $\eta_\mathrm{ret}=\eta_\mathrm{tot}/\eta_\mathrm{store}$ is significantly larger than the storage efficiency~$\eta_\mathrm{store}$, since the total efficiency $\eta_\mathrm{tot}$ exceeds $\eta_\mathrm{store}^2$. This indicates that the signal pulse shape is not optimal --- an ideal memory has equal storage and retrieval efficiencies~\cite{Gorshkov_2007_PRAMOP_Photon-storage-in-Lambda-type-}. Correctly mode matching the signal pulse profile to the control should significantly increase the efficiency.

Part (c) of figure~\ref{fig:eff_error} shows an extrapolation of the theoretical prediction for $\eta_\mathrm{tot}$ to larger write/read pulse energies. This indicates that $\eta_\mathrm{tot}\sim 30\%$ is achievable for pulse energies $\sim 15$ nJ. Also plotted is the optimal attainable efficiency in the present configuration with retrieval in the forward direction, along with the optimal efficiency for backward retrieval \cite{Gorshkov_2007_PRAMOP_Photon-storage-in-Lambda-type-,Surmacz_2008_PRAMOP_Efficient-spatially-resolved-m}. To achieve these bounds the signal field requires appropriate shaping --- the distortion due to the etalons should be compensated, along with the dynamic Stark shift due to the strong write field. Re-absorption of the signal limits the efficiency to around $60\%$ for forward retrieval, but efficiencies above $90\%$ can be reached employing phase-matched backward retrieval \cite{Surmacz_2008_PRAMOP_Efficient-spatially-resolved-m}.
\begin{figure}[h]
\begin{center}
\centerline{\includegraphics[width=9cm]{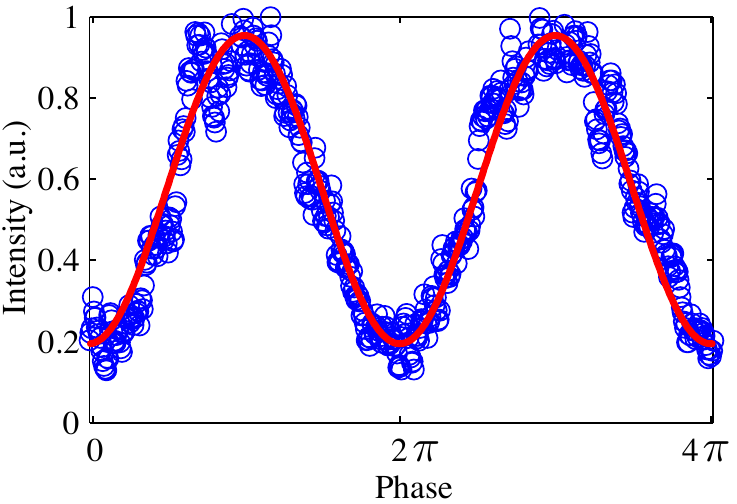}}
\caption{Raw interference and fit of stored and retrieved signal. Circles indicate experimental data, the red curve is a least squares fit.  
A linear scan of the path length difference in the interferometer results in sinusoidal oscillations of the total intensity. The high visibility  of $86\%$ (normalized for interferometer instability) of the interference demonstrates the coherence of the memory, and matches with theoretical calculations, suggesting that the memory interaction is perfectly coherent.}
\label{fig:interference}
\end{center}
\end{figure}

Because we retrieve the stored signal after just 12.5 ns (the round-trip time of our oscillator), the efficiency we observe is not limited by decoherence, which is only significant over much longer timescales. Instead, it is a direct probe of the intrinsic efficiency of the Raman memory interaction. 
In addition, it is easy in this configuration to delay a copy of the signal pulse and interfere it directly with the retrieved pulse, to demonstrate the coherence of the interaction (Fig.~\ref{fig:interference}).
The fringe visibility of $86\pm 5\%$ indicates that the memory is highly coherent. In fact, the theoretical model (supplementary information) predicts a distortion of the retrieved field due to dispersion and Stark shifts (these can be eliminated with backward retrieval), yielding a maximum visibility of $83\%$. This suggests that the memory interaction is perfectly coherent.

In summary, we have demonstrated a broadband quantum-capable memory by coherently storing and retrieving signal pulses with bandwidths greater than 1 GHz. This is an increase of almost a factor of 1000 compared to existing quantum memories.  Storage efficiencies up to 30\% and retrieval efficiencies as high as 50\% were observed, and increasing the power of the control pulses should allow further improvements. The excellent coherence of the memory was directly verified by interfering the stored and the retrieved pulses. 
Such high-speed memories  will form the basis of fast, controllable and robust photonic quantum information processors in the near-future.

\setcounter{figure}{0}
\setcounter{equation}{0}
\renewcommand{\thefigure}{S\arabic{figure}}
\renewcommand{\theequation}{S\arabic{equation}}

\section*{Methods}

\begin{figure}[h]
\begin{center}
\includegraphics[width = 11cm]{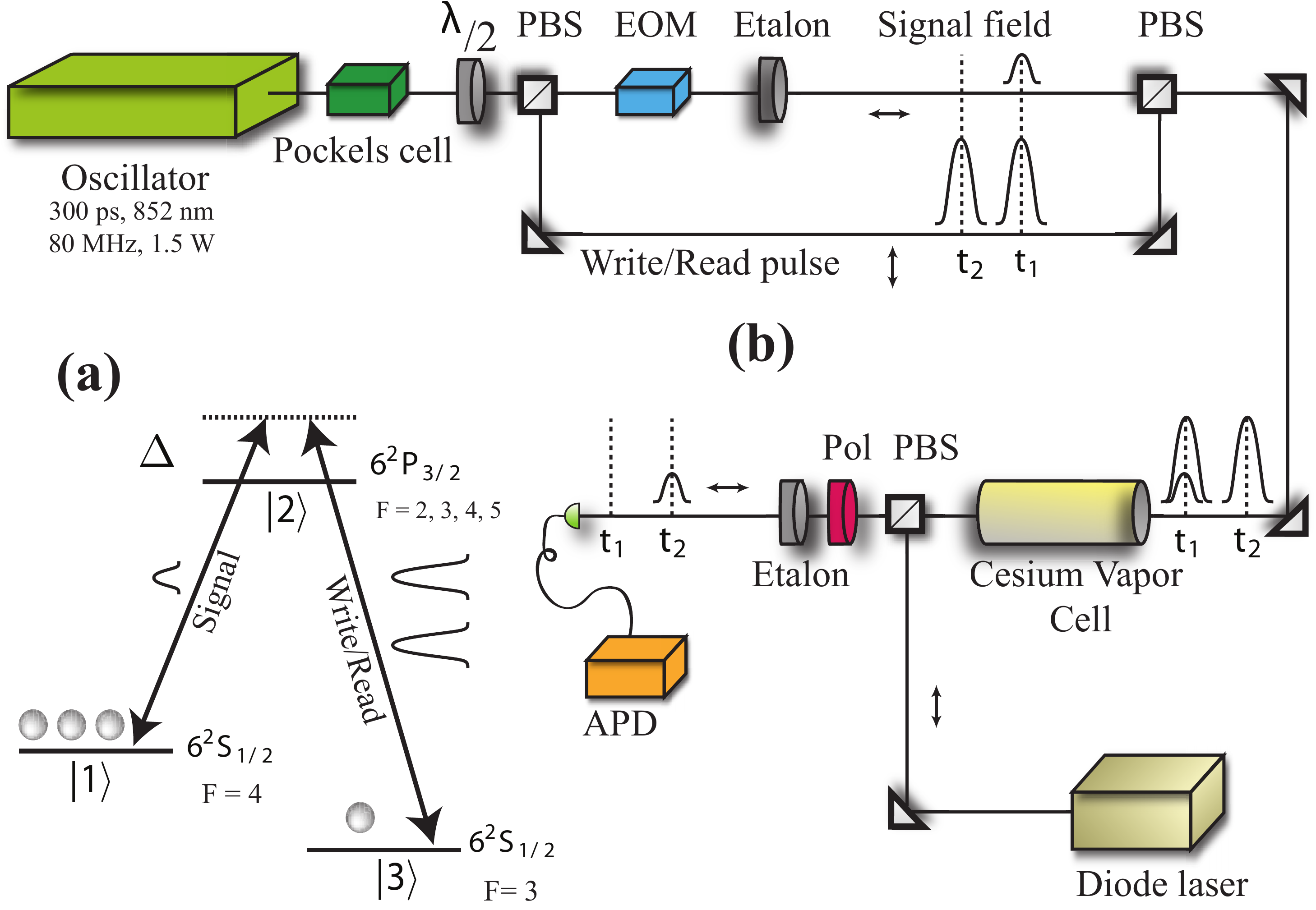}
\caption{(a) $\Lambda$-level scheme for Raman memory. (b) Experimental setup. A cesium vapor cell is optically prepared with a diode laser. In time bin $t_1$ an incoming signal pulse is mapped by a strong write pulse into a spin wave excitation in the atomic cesium ensemble. At $t_2$ a strong read pulse reconverts the excitation into a light pulse. After polarization filtering (Pol), the retrieved signal is detected by a high speed avalanche photo detector (APD). Vertical polarization is indicated as ($\updownarrow$) and  horizontal polarization as ($\leftrightarrow$). $\Delta$ is the detuning from the atomic resonance.}
\label{fig:experiment}
\end{center}
\end{figure}

The atomic states of cesium involved in the Raman protocol are shown in part (a) of Figure \ref{fig:experiment}(supplementary information). The experimental layout is described in part (b). The read, write and signal pulses are derived from a Ti:Sapph oscillator and have a FWHM of 300 ps (1.5 GHz bandwidth). The fundamental Ti:Sapph laser frequency is tuned 18.4 GHz to the blue of the $\ket{2}$-$\ket{3}$   transition in Figure \ref{fig:experiment}  (a). A Pockels cell selects two consecutive pulses, separated by 12.5 ns. The laser beam is split into a strong control arm with vertical polarization ($\updownarrow$)   and a very weak signal arm with horizontal polarization ($\leftrightarrow$) . The control arm is delayed by 12.5 ns with respect to the signal arm such that the first pulse in the control arm overlaps in time with the last pulse in the signal arm. An electro-optic modulator (EOM) is used in the signal arm to generate sidebands 9.2 GHz shifted from the fundamental laser frequency. After spectral filtering with Fabry Perot etalons only the 9.2 GHz red-shifted sideband corresponding to the $\ket{1}$-$\ket{2}$   transition is transmitted and used as the signal field. The control and the signal beam, spectrally separated by 9.2 GHz, are recombined and made collinear. They are focused with a beam waist of 350 $\mu$m into the 7 cm long vapor cell filled with cesium and 20 torr of neon buffer gas. Polarization- and spectral filtering are used after the cell to extinguish the strong write and read pulses and transmit the signal field only. A high-speed avalanche photo detector with a bandwidth of 1 GHz detects the very weak signal pulse. The atomic ensemble is initially prepared in the ground state $\ket{1}$   by optical pumping using an external cavity diode laser tuned to resonance with the  $\ket{2}$-$\ket{3}$   transition. The underlying theory for the Raman memory scheme is derived in \cite{Gorshkov_2007_PRAMOP_Photon-storage-in-Lambda-type-,Nunn_2007_PRAMOP_Mapping-broadband-single-photo} and briefly summarized in the supplementary section.

To investigate the coherence properties of the memory, a copy of the incident signal field is attenuated, delayed and overlapped with the retrieved signal in a Mach-Zehnder configuration. We correct for imperfections in the interferometer by interfering the signal and a replica as a benchmark; this yields a visibility of $86\pm 5\%$ ($67\%$ uncorrected) for the memory.

\section*{Acknowledgements}
We thank D. Stacey, P. Walther and M. G. Raymer  for useful discussions. This work was supported by the
EPSRC through the QIP IRC (GR/S82716/01) and project EP/C51933/01. KFR and VOL
were supported by the Marie-Curie-Network EMALI. BJS gratefully acknowledges support
from the Natural Sciences and Engineering Research Council of Canada and from the
Royal Society. IAW was supported in part by the European Commission under the Integrated
Project Qubit Applications (QAP) funded by the IST directorate as Contract Number
015848, and the Royal Society.

\section*{Author contribution}
All authors contributed to this work, and agree to the contents of the paper.

Correspondence and requests for materials should be addressed to I. A. Walmsley (i.walmsley1@physics.ox.ac.uk).
\section*{Supplementary information}

\subsection*{Experimental details}

\subsection*{Theory}

The storage and retrieval efficiencies are calculated by solving the semi-classical linearized Maxwell-Bloch equations for the system. The signal field, with amplitude $A$, propagates through an ensemble of $\Lambda$-level atoms in the presence of the write field, whose temporal profile is described by the time-dependent Rabi frequency $\Omega(\tau)$ \cite{Gorshkov_2007_PRAMOP_Photon-storage-in-Lambda-type-,Nunn_2007_PRAMOP_Mapping-broadband-single-photo}. Provided that the detuning $\Delta$ is the dominant frequency in the problem, the optical polarization can be adiabatically eliminated, yielding an explicit expression for the spin wave amplitude $B$ at the end of the storage interaction. The spin wave can be expressed as \cite{Gorshkov_2007_PRAMOP_Photon-storage-in-Lambda-type-,Nunn_2008_PRL_Multimode-Memories-in-Atomic-E},
\be
B_\mathrm{mem}(z) = \int_{-\infty} ^{\infty} f(\tau) J_0\left[2C\sqrt{(1-\omega(\tau)) z}\right] A_\mathrm{in}(\tau)\,\mathrm{d}\tau,
\label{eq:spinwave}
\ee
where $A_\mathrm{in}$ is the amplitude of the incident signal field to be stored, $J_0$ is a Bessel function, and the number $C^2=d\gamma W/\Delta^2$ quantifies the Raman memory coupling, with $d$ the resonant optical depth \cite{Gorshkov_2007_PRAMOP_Photon-storage-in-Lambda-type-} and $\gamma$ the homogeneous linewidth of the excited state $\ket{2}$. Here we introduced the dimensionless integrated Rabi frequency $\omega(\tau)=\frac{1}{W}\int_{-\infty}^\tau \left|\Omega(\tau')\right|^2\,\mathrm{d}\tau'$, and the normalized Stark-shifted Rabi frequency $f(\tau)=C e^{\mathrm{i}W\omega(\tau)/\Delta}\Omega(\tau)/\sqrt{W}$. The constant $W$, proportional to the control pulse energy, is defined so that $\omega(\infty)=1$, while the longitudinal coordinate $z$ is normalized so that $z=1$ represents the exit face of the ensemble.

The storage efficiency $\eta_\mathrm{store}=N_\mathrm{mem}/N_\mathrm{in}$ is the ratio of the number $N_\mathrm{mem}=\int_0^1 \left|B_\mathrm{mem}(z)\right|^2\,\mathrm{d}z$ of final spin wave excitations to the number $N_\mathrm{in}=\int_{-\infty}^\infty \left|A_\mathrm{in}(\tau)\right|^2\,\mathrm{d}\tau$ of incident signal photons. If upon retrieval $N_\mathrm{out}=\int_{-\infty}^\infty \left|A_\mathrm{out}(\tau)\right|^2\,\mathrm{d}\tau$ photons are recovered from the memory, the total efficiency $\eta_\mathrm{tot}=N_\mathrm{out}/N_\mathrm{in}$ can be computed using the formula
\begin{equation}
\label{A_out}
A_\mathrm{out}(\tau) = f^*(\tau) \int_0^1 J_0\left[2C\sqrt{\omega(\tau)(1- z)}\right]B_\mathrm{mem}(z)\,\mathrm{d}z
\end{equation}
for the retrieved signal field. In general, the field $A_\mathrm{out}$ is different to $A_\mathrm{in}$ --- physically these differences originate in the dispersion associated with propagation through the ensemble, and the Stark shift due to the time-varying control field. The visibility of interference between the incident and retrieved signals is therefore smaller than $1$, even for a perfectly coherent memory. If optimal storage and backward retrieval is possible, these distortions can, however, be eliminated \cite{Surmacz_2008_PRAMOP_Efficient-spatially-resolved-m}.

\bibliographystyle{unsrt}

\begin{thebibliography}{10}
\bibitem{Kok_2007_RMP_Linear-optical-quantum-computi}
P.~Kok, WJ~Munro, K.~Nemoto, TC~Ralph, J.P. Dowling, and GJ~Milburn.
\newblock {Linear optical quantum computing with photonic qubits}.
\newblock {\em Reviews of Modern Physics}, 79(1):135--174, 2007.

\bibitem{Duan_2001_N_Long-distance-quantum-communic}
L.-M. Duan, M.~D. Lukin, J.~I. Cirac, and P.~Zoller.
\newblock Long-distance quantum communication with atomic ensembles and linear
  optics.
\newblock {\em Nature}, 414(6862):413--418, November 2001.

\bibitem{Hetet_2008_PRL_Electro-Optic-Quantum-Memory-f}
G.~Hetet, J.~J. Longdell, A.~L. Alexander, P.~K. Lam, and M.~J. Sellars.
\newblock {{Electro-Optic} Quantum Memory for Light Using {Two-Level} Atoms}.
\newblock {\em Physical Review Letters}, 100(2):023601--4, 2008.

\bibitem{Riedmatten_2008_N_A-solid-state-light-matter-int}
Hugues de~Riedmatten, Mikael Afzelius, Matthias~U. Staudt, Christoph Simon, and
  Nicolas Gisin.
\newblock {A solid-state light-matter interface at the single-photon level}.
\newblock {\em Nature}, 456(7223):773--777, December 2008.

\bibitem{Schnorrberger_2009_PRL_Electromagnetically-Induced-Tr}
U.~Schnorrberger, J.~D. Thompson, S.~Trotzky, R.~Pugatch, N.~Davidson, S.~Kuhr,
  and I.~Bloch.
\newblock {Electromagnetically Induced Transparency and Light Storage in an
  Atomic Mott Insulator}.
\newblock {\em Physical Review Letters}, 103(3):033003--4, July 2009.

\bibitem{Boozer_2007_PRL_Reversible-State-Transfer-betw}
A.~D. Boozer, A.~Boca, R.~Miller, T.~E. Northup, and H.~J. Kimble.
\newblock {Reversible State Transfer between Light and a Single Trapped Atom}.
\newblock {\em Physical Review Letters}, 98(19):193601--4, May 2007.

\bibitem{Gorshkov_2007_PRAMOP_Photon-storage-in-Lambda-type-}
Alexey~V. Gorshkov, Axel Andre, Mikhail~D. Lukin, and Anders~S. Sorensen.
\newblock {Photon storage in Lambda-type optically dense atomic media. {II.}
  Free-space model}.
\newblock {\em Physical Review A {(Atomic,} Molecular, and Optical Physics)},
  76(3):033805--25, 2007.

\bibitem{Nunn_2007_PRAMOP_Mapping-broadband-single-photo}
J.~Nunn, I.~A. Walmsley, M.~G. Raymer, K.~Surmacz, F.~C. Waldermann, Z.~Wang,
  and D.~Jaksch.
\newblock {Mapping broadband single-photon wave packets into an atomic memory}.
\newblock {\em Physical Review A {(Atomic,} Molecular, and Optical Physics)},
  75(1):011401--4, 2007.

\bibitem{Camacho_2009_P_Four-wave-mixing-stopped-light}
Ryan~M. Camacho, Praveen~K. Vudyasetu, and John~C. Howell.
\newblock {Four-wave-mixing stopped light in hot atomic rubidium vapour}.
\newblock {\em Nat Photon}, 3(2):103--106, February 2009.

\bibitem{Shor_1997_SJSSC_Polynomial-Time-Algorithms-for}
Peter Shor.
\newblock Polynomial-time algorithms for prime factorization and discrete
  logarithms on a quantum computer.
\newblock {\em SIAM J. Sci. Statist. Comput.}, 26:1484, 1997.

\bibitem{Sangouard_2009_EA0_Quantum-repeaters-based-on-ato}
N.~Sangouard, C.~Simon, H.~de~Riedmatten, and N.~Gisin.
\newblock {Quantum repeaters based on atomic ensembles and linear optics}.
\newblock {\em eprint arXiv: 0906.2699}, 2009.

\bibitem{Barrett_2008__Scalable-quantum-computing-wit}
S.~D. Barrett, P.~P. Rohde, and T.~M. Stace.
\newblock {Scalable quantum computing with atomic ensembles}, 2008.

\bibitem{Choi_2008_N_Mapping-photonic-entanglement-}
K.~S. Choi, H.~Deng, J.~Laurat, and H.~J. Kimble.
\newblock {Mapping photonic entanglement into and out of a quantum memory}.
\newblock {\em Nature}, 452(7183):67--71, March 2008.

\bibitem{Nunn_2008_PRL_Multimode-Memories-in-Atomic-E}
J.~Nunn, K.~Reim, K.~C. Lee, V.~O. Lorenz, B.~J. Sussman, I.~A. Walmsley, and
  D.~Jaksch.
\newblock {Multimode Memories in Atomic Ensembles}.
\newblock {\em Physical Review Letters}, 101(26):260502--4, December 2008.

\bibitem{Simon_2007_PRL_Quantum-Repeaters-with-Photon-}
Christoph Simon, Hugues de~Riedmatten, Mikael Afzelius, Nicolas Sangouard, Hugo
  Zbinden, and Nicolas Gisin.
\newblock Quantum repeaters with photon pair sources and multimode memories.
\newblock {\em Physical Review Letters}, 98(19):190503, 2007.

\bibitem{Eisaman_2005_N_Electromagnetically-induced-tr}
M.~D. Eisaman, A.~Andre, F.~Massou, M.~Fleischhauer, A.~S. Zibrov, and M.~D.
  Lukin.
\newblock Electromagnetically induced transparency with tunable single-photon
  pulses.
\newblock {\em Nature}, 438(7069):837--841, December 2005.

\bibitem{Harris_1997_PT_Electromagnetically-Induced-Tr}
Stephen~E. Harris.
\newblock {Electromagnetically Induced Transparency}.
\newblock {\em Physics Today}, 50(7):36, 1997.

\bibitem{Liu_2001_N_Observation-of-coherent-optica}
Chien Liu, Zachary Dutton, Cyrus~H. Behroozi, and Lene~Vestergaard Hau.
\newblock {Observation of coherent optical information storage in an atomic
  medium using halted light pulses}.
\newblock {\em Nature}, 409(6819):490--493, 2001.

\bibitem{Alexander_2006_PRL_Photon-Echoes-Produced-by-Swit}
A.~L. Alexander, J.~J. Longdell, M.~J. Sellars, and N.~B. Manson.
\newblock {Photon Echoes Produced by Switching Electric Fields}.
\newblock {\em Physical Review Letters}, 96(4):043602--4, February 2006.

\bibitem{Kraus_2006_PRAMOP_Quantum-memory-for-nonstationa}
B.~Kraus, W.~Tittel, N.~Gisin, M.~Nilsson, S.~Kroll, and J.~I. Cirac.
\newblock {Quantum memory for nonstationary light fields based on controlled
  reversible inhomogeneous broadening}.
\newblock {\em Physical Review A {(Atomic,} Molecular, and Optical Physics)},
  73(2):020302--4, February 2006.

\bibitem{Staudt_2007_PRL_Fidelity-of-an-Optical-Memory-}
M.~U. Staudt, S.~R. {Hastings-Simon}, M.~Nilsson, M.~Afzelius, V.~Scarani,
  R.~Ricken, H.~Suche, W.~Sohler, W.~Tittel, and N.~Gisin.
\newblock {Fidelity of an Optical Memory Based on Stimulated Photon Echoes}.
\newblock {\em Physical Review Letters}, 98(11):113601--4, March 2007.

\bibitem{Chaneliere_2009_0_Efficient-light-storage-in-a-c}
T.~Chaneli{\`e}re, M.~Afzelius, and J-L.~Le Gou{\"e}t.
\newblock {Efficient light storage in a crystal using an Atomic Frequency
  Comb}.
\newblock {\em arXiv:0902.2048}, February 2009.

\bibitem{Zhu_2007_S_Stored-light-in-an-optical-fib}
Z.~Zhu, D.J. Gauthier, and R.W. Boyd.
\newblock {Stored light in an optical fiber via stimulated Brillouin
  scattering}.
\newblock {\em Science}, 318(5857):1748, 2007.

\bibitem{Hosseini_2009_N_Coherent-optical-pulse-sequenc}
Mahdi Hosseini, Ben~M. Sparkes, Gabriel Hetet, Jevon~J. Longdell, Ping~Koy Lam,
  and Ben~C. Buchler.
\newblock {Coherent optical pulse sequencer for quantum applications}.
\newblock {\em Nature}, 461(7261):241--245, 2009.

\bibitem{Novikova_2008_PRAMOP_Optimal-light-storage-with-ful}
Irina Novikova, Nathaniel~B. Phillips, and Alexey~V. Gorshkov.
\newblock {Optimal light storage with full pulse-shape control}.
\newblock {\em Physical Review A {(Atomic,} Molecular, and Optical Physics)},
  78(2):021802--4, 2008.

\bibitem{Longdell_2005_PRL_Stopped-Light-with-Storage-Tim}
J.~J. Longdell, E.~Fraval, M.~J. Sellars, and N.~B. Manson.
\newblock {Stopped Light with Storage Times Greater than One Second Using
  Electromagnetically Induced Transparency in a Solid}.
\newblock {\em Physical Review Letters}, 95(6):063601, 2005.
\newblock Copyright {(C)} 2009 The American Physical Society; Please report any
  problems to prola@aps.org.

\bibitem{Kozhekin_2000_PR_Quantum-memory-for-light}
A.~E. Kozhekin, K.~M�lmer, and E.~Polzik.
\newblock {Quantum memory for light}.
\newblock {\em Physical Review A}, 62(3):033809, 2000.
\newblock Copyright {(C)} 2009 The American Physical Society; Please report any
  problems to prola@aps.org.

\bibitem{Surmacz_2008_PRAMOP_Efficient-spatially-resolved-m}
K.~Surmacz, J.~Nunn, K.~Reim, K.~C. Lee, V.~O. Lorenz, B.~Sussman, I.~A.
  Walmsley, and D.~Jaksch.
\newblock {Efficient spatially resolved multimode quantum memory}.
\newblock {\em Physical Review A {(Atomic,} Molecular, and Optical Physics)},
  78(3):033806--9, 2008.

\end{thebibliography}

\end{document}